\documentclass[draftcls,onecolumn]{IEEEtran}

\usepackage{amsmath,graphicx,subfigure,booktabs,colortbl,amsfonts}
\usepackage{amssymb,algorithm,algpseudocode}
\usepackage{mathrsfs,url,bm,bbm}
\newtheorem{Rem}{Remark}

\begin{document}

\title{Multi-Sensor Control for Multi-Object Bayes Filters}
\author{Xiaoying~Wang,
        Reza~Hoseinnezhad$^*$,
        Amirali~K.~Gostar, 
        Tharindu~Rathnayake, Benlian Xu and
        Alireza Bab-Hadiashar
\thanks{$^*$ Corresponding Author; email: rezah@rmit.edu.au}
\thanks{X.~Wang and B.~Xu are with Changshu Institute of Technology, Changshu, Soochow, P. R. of China. X.~Wang is also a visiting researcher at School of Engineering, RMIT University, Victoria, Australia.}
\thanks{R.~Hoseinnezhad, A.~K.~Gostar, T.~Rathnayake and A.~Bab-Hadiashar are with School of Engineering, RMIT University, Victoria, Australia.}
\thanks{Manuscript received mmmm dd, yyyy.}}

\markboth{XXXXXX,~Vol.~xx, No.~xx, mmmm~yyyy}%
{Wang~\MakeLowercase{\textit{et al.}}: Multi-Sensor Control for Multi-Object Bayes Filters}

\maketitle

\begin{abstract}
Sensor management in multi-object stochastic systems is a theoretically and computationally challenging problem. This paper presents a novel approach to the multi-target multi-sensor control problem within the partially observed Markov decision process (POMDP) framework. We model the multi-object state as a labeled multi-Bernoulli random finite set (RFS), and use the labeled multi-Bernoulli filter in conjunction with minimizing a task-driven control objective function: posterior expected error of cardinality and state (PEECS). A major contribution is a guided search for multi-dimensional optimization in the multi-sensor control command space, using coordinate descent method. In conjunction with the Generalized Covariance Intersection  method for multi-sensor fusion, a fast multi-sensor algorithm is achieved. Numerical studies are presented in several scenarios where numerous controllable (mobile) sensors track multiple moving targets with different levels of observability. The results show that our method works significantly faster than the approach taken by a state of art method, with similar tracking errors.
\end{abstract}

\begin{IEEEkeywords}
partially observed Markov decision process, multi-target tracking, random finite sets, labeled multi-Bernoulli filter, coordinate descent.
\end{IEEEkeywords}

\section{Introduction}
Multi-object multi-sensor management/control is a challenging optimal nonlinear control problem focused on directing multiple sensors to obtain \textit{most informative} measurements for the purpose of multi-object filtering~\cite{Mahler_Zajic}. This problem is different from classical control problems as the overall controlled system is a highly complex stochastic multi-object system, where not only the number of objects vary randomly in time, but also the measurements returned by each sensor are subject to missed detections and false alarms. Indeed, the multi-object state and multi-object observations are inherently
finite-set-valued, and standard optimal control techniques are not directly applicable.

In stochastic multi-object systems, we can still cast the multi-object multi-sensor control problem as a partially observed Markov decision process (POMDP), where the states and observations are instead
finite-set-valued, and control vectors are drawn from a set of admissible sensor actions based on the current information states, which are then assessed against the values of an objective function associated with each multi-sensor action~\cite{Castanon_2008}. In this framework, a solution would include three major steps: (1) modeling the overall system as a stochastic multi-object system, (2) devising a tractable (accurate or approximate) way to propagate the multi-object posterior, and (3) solving an optimization problem to find the multi-sensor control command, according to an objective function. This paper presents a formulation of the multi-sensor control problem as a POMDP with finite-set-valued states and measurements, a labeled random set filter used to propagate the multi-object posterior, and a task-driven objective (cost) function.

To our knowledge, the problem of multi-sensor control for labeled random set filters is only recently considered by Meng~\textit{et~al.}~\cite{meng_CS_2016}. In this method, local Vo-Vo filters\footnote{The authors of~\cite{vo_glmb_TSP_2014} originally called their  filter the delta-Generalized Multi-Bernoulli ($\delta-$GLMB) filter. In this work, we follow the simpler name suggested by R.~Mahler in his book~\cite{book_Mahler_2014}.} are operating at each sensor node, and the resulting Vo-Vo densities (posteriors) are fused using the Generalized Covariance Intersection (GCI) rule as formulated in~\cite{consensus_labelled_RFS_2016}. The approach opted by Meng~\textit{et~al.}~\cite{meng_CS_2016} to solve the multi-sensor control problem is an \textit{exhaustive search} scheme, in which the objective function is computed for all possible combinations of sensor control actions. This approach works well for a few sensors only, but in presence of numerous sensors, may become computationally intractable. 

The major contribution of this paper is the introduction of a guided search to solve the multi-dimensional discrete optimization problem embedded in multi-sensor control. We avoid the curse of dimensionality by using an accelerated scheme inspired by the coordinate descent method~\cite{coordinate_descent_Tseng_2001}.   This leads to significant improvement in the runtime of the algorithm and its real-time feasibility, especially in presence of numerous sensors. Another contribution is the detailed sequential Monte-Carlo (SMC) implementation of the proposed multi-sensor control framework with Labeled Multi-Bernoulli~(LMB) filters running in each sensor node. The novel idea inherent in the proposed SMC implementation is that sensor control and the actual filters are all implemented using the same particles, hence substantial savings are achieved in terms of memory and computational requirements. We also experimentally analyse the computational complexity of the proposed method and demonstrate that it varies almost quadratically with the number of controlled sensors (polynomial complexity). This is while an exhaustive search similar to the one used in~\cite{meng_CS_2016} has exponential (hence, non-polynomial) complexity.

Extensive simulation studies involving numerous controllable sensors demonstrate that our method returns acceptable tracking results quantified in terms of OSPA error values~\cite{vo_OSPA_TSP_2008}. Indeed, in comparison to the state of art (running exhaustive search in an approach similar to~\cite{meng_CS_2016}), the proposed multi-sensor control method returns similar tracking errors but converges significantly faster. 

The organization of the paper is as follows. Section~\ref{sec:prob_stat} presents a formalized statement of the multi-sensor control problem in POMDP framework and sets out the background and design requirements for various components of the framework. The proposed multi-sensor control solution is then presented in section~\ref{sec:approach}, outlining the general framework and proposed choices for its components, as well as a step-by-step algorithm for the SMC implementation. Simulation results are presented in section~\ref{sec:sim_res}. Section~\ref{sec:conc} concludes the paper.

\section{Problem Statement}
\label{sec:prob_stat}

Consider a stochastic multi-object system, in which at any discrete (or sampling) time $k$, the multi-object state $\bm{X}_k$ is a labeled random finite set (RFS) comprised of a random number of single-object states,
\begin{equation}
\bm{X}_k = \left\{
(x_{1,k},\ell_1), \ldots, (x_{n_k,k},\ell_{n_k})
\right\} \in \mathcal{F}(\mathbb{X}\times\mathbb{L})
\end{equation}
where $\mathbb{X}$ and $\mathbb{L}$ denote the state and label spaces, respectively, and $\mathcal{F}(\cdot)$ means ``all finite subsets of.''

The system is modeled as a one-step-ahead Markovian process which is characterized by a transition density $\bm{f}(\bm{X}_k|\bm{X}_{k-1})$. A practical approximation for the process can be formulated based on assuming that while transiting from time $k-1$ to time $k$, each existing object $\bm{x}$ independently continues to exist with a survival probability $p_S(\bm{x})$ and single-object transition density $f_{k|k-1}(\cdot|\bm{x})$, and a number of new objects are born according to a given RFS density.

At each time $k$, the multi-object state is partially observed by a network of $n_s$ sensors, each returning a set of measurements (called detections or point measurements). Let $Z_i$ be the measurement set returned by the $i$-th sensor, $s_i,~(i=1:n_s)$. Denoting the space of point measurements by $\mathbb{Z}$, the space of measurement sets will be $\mathbbm{Z} = \mathcal{F}(\mathbb{Z})$. Each sensor $s_i$ can be controlled (e.g. translated, rotated) according to a sensor command $u_i \in \mathbb{U}$ where $\mathbb{U}$ is a finite space of sensor commands. The cumulated measurement is an $n_s$-tuple of measurement sets,
\begin{equation}
\mathfrak{Z}_k = (Z_1,\ldots,Z_{n_s}) \in \mathbbm{Z}^{n_s}.
\end{equation} 
The relationship between the multi-sensor measurement and the multi-object state is stochastically modeled by the multi-object likelihood function $\bm{g}(\mathfrak{Z}_k|\bm{X}_k,\mathfrak{u})$, where 
$$\mathfrak{u} = (u_1,\cdots,u_{n_s}) \in \mathbb{U}^{n_s}$$ 
is the multi-sensor command. The likelihood function is usually modeled in terms of a single-object likelihood $g(z|\bm{x},u)$, a state-dependent detection probability $p_D(\bm{x})$ and assuming a Poisson process for the number of false alarms which together are modeled as a Poisson RFS characterized by an intensity function $\kappa(\cdot)$.

The multi-sensor control problem can be formally cast in the framework of the following 6-tuple discrete-time POMDP:
\begin{equation}
\Psi = \left\{
\mathbbm{X},
\mathbb{U}^{n_s},
\bm{f}(\cdot|\cdot),
\mathbbm{Z}^{n_s},
\bm{g}(\cdot|\cdot,\mathfrak{u}),
\nu(\mathfrak{u};\cdot)
\right\}
\label{eq:pomdp1}
\end{equation}
where $\nu(\mathfrak{u}; \cdot)$ is an objective function that associates a reward or cost with a choice of multi-sensor control command $\mathfrak{u} = (u_1,\ldots,u_{n_s})$ given the recent multi-object state $\bm{X}_{k-1}$ or its statistical characteristics. In a one-step-ahead multi-sensor control solution, the aim is to find the multi-sensor command, 
$$
\mathfrak{u}^* = (u^*_1, \ldots, u^*_{n_s})
$$
that satisfies
\begin{equation}
(u^*_1,\ldots,u^*_{n_s}) = \underset{{(u_1,\ldots,u_{n_s}) \in \mathbb{U}^{n_s}}}{\arg\min\!/\!\max} ~~~\nu(u_1,\ldots,u_{n_s};\bm{X}_{k-1}).
\end{equation}

\subsection{Multi-target Bayes filter within POMDP}

Given the POMDP with the components given in~\eqref{eq:pomdp1}, the probability density of multi-object state of the system can be recursively estimated by a multi-target Bayes filter. Let us denote the density of multi-object state at time $k$ by $\pi_k(\bm{X}_k|\mathfrak{Z}_{1:k})$, where $\mathfrak{Z}_{1:k}$ denotes the ensemble of all multi-sensor measurements accumulated up to time $k$. In a Bayesian filtering scheme, the density is recursively propagated through two steps: prediction and update~\cite{vo_glmb_TSP_2014,book_Mahler_2014}. The predicted density is computed by the multi-object Chapman-Kolmogorov equation:
\begin{equation}
\begin{array}{l}
\pi_{k|k-1}(\bm{X}_k|\mathfrak{Z}_{1:({k-1})}) = \\
\hspace{1.0cm}
\int \pi_{k-1}(\bm{X}_{k-1}|\mathfrak{Z}_{1:(k-1)}) \bm{f}(\bm{X}_k|\bm{X}_{k-1}) \delta \bm{X}_{k-1}.
\end{array}
\label{eq:bayesian_pred_general}
\end{equation}
With the arrival of new observations $\mathfrak{Z}_k = (Z_1,\ldots,Z_{n_s})$ from the sensors controlled by a multi-sensor action $\mathfrak{u}_k$, a posterior density is obtained using multi-object Bayes' rule:
\begin{equation}
\pi_k(\bm{X}_k|\mathfrak{Z}_{1:k}) = \frac{\bm{g}(\mathfrak{Z}_k|\bm{X}_k,\mathfrak{u}_k)\ \pi_{k|k-1}(X_{k})}{\int \bm{g}(\mathfrak{Z}_k|\bm{X},\mathfrak{u}_k)\ \pi_{k|k-1}(X)\ \delta\bm{X}}.
\label{eq:bayesian_update_general}
\end{equation}

\begin{Rem}
	Given the posterior recursion~\eqref{eq:bayesian_pred_general}~and \eqref{eq:bayesian_update_general}, the objective function component of the POMDP in~\eqref{eq:pomdp1}, is usually defined as a function of the probability density of the multi-object state, and the optimization component of the multi-sensor control framework is expressed as
	\begin{equation}
	\mathfrak{u}^* = \underset{{\mathfrak{u} \in \mathbb{U}^{n_s}}}{\arg\min\!/\!\max} ~~~\nu(\mathfrak{u};\pi_{k-1}(\bm{X}_{k-1}).
	\end{equation}
\end{Rem}
\begin{Rem} 
The integrals in \eqref{eq:bayesian_pred_general} and \eqref{eq:bayesian_update_general} are set integrals as defined in~\cite{book_Mahler_2014}. The recursion \eqref{eq:bayesian_pred_general} and \eqref{eq:bayesian_update_general} has no analytic solution in general. An SMC implementation of the Bayes multi-object filter (with RFS states without labels) is given in~\cite{vo_montecarlo_aes2005}. However, this technique is computationally prohibitive which at best is able to accommodate a small
number of targets. This SMC implementation of the multi-object Bayes filter was employed by the multi-target sensor control algorithm proposed in~\cite{ristic_automatica_2010}.
\end{Rem}

Due to general intractability of propagation of the full posterior density given by \eqref{eq:bayesian_pred_general} and \eqref{eq:bayesian_update_general}, several alternatives have been proposed which are designed to propagate important statistics or parameters instead of the full posterior. Well-known examples of such filters are probability hypothesis density (PHD) filter and its cardinalized version (CPHD)~\cite{book_Mahler_2014}, and the multi-Bernoulli filter and its cardinality-balanced version (CB-MeMBer)~\cite{vo_cbmember_TSP_2009}. In a series of works~\cite{vo_tbd_TSP_2010,our_Fusion_12,our_TSP_visual_tracking,our_icassp_2010,our_pattern_rec_2012,robust_MB_JSTSP}, various implementations of these filters such as SMC and track-before-detect (TBD) were introduced, as well as a robust version of multi-Bernoulli filter. These methods  can not generate target tracks (using labels) in a  rigorously mathematical way, and are usually applied in conjunction with a label management strategy~\cite{our_pattern_rec_2012,our_TSP_visual_tracking}. 

Since 2010, a series of random set filters have been developed, in which the multi-object random state includes label. The \textit{labeled random finite sets} were shown to admit conjugacy of a particular form of prior density (the Vo-Vo density) with the general multiple point measurement set likelihood~\cite{vo_glmb_TSP_2013}. Following this result, the Vo-Vo filter was introduced~\cite{vo_glmb_TSP_2014,vo_glmb_TSP_2015}. Variants of the Vo-Vo filter such as the labeled multi-Bernoulli (LMB) filter~\cite{vo_lmb_TSP_2014} and M-$\delta$-GLMB filter~\cite{m_delta_glmb_SPL_2016} were also proposed and applied in various applications. 

The proposed multi-sensor control framework can be implemented with different multi-object filters. For the sake of completion and presenting a step-by-step pseudocode, we have chosen to implement our method with the LMB filter.

\subsection{Objective function}
The choice of objective function $\nu(u;\pi_{k-1}(\cdot))$ is a critical part of the control solution design task. The objective functions commonly used in sensor control solutions in the stochastic signal processing and control literature, can be generally divided into two types: information-driven and task-driven. The information-driven reward function quantifies the expected information gain from prior to posterior after a hypothesized sensor control action. For example, R\'enyi divergence was usedby Ristic \textit{et~al.}~\cite{ristic_automatica_2010,ristic_reward_PHD_TAES_2011} for sensor control with random set filters in general~\cite{ristic_automatica_2010} and PHD filters in particular~\cite{ristic_reward_PHD_TAES_2011}. Recently, in a number of works, the Cauchy-Schwarz divergence has been adopted as the reward function~\cite{vo_CS_Poisson,hoang_CS_2014,vo_CS_fusion2015}.

The task-driven cost functions are usually formulated in terms of the expected error of estimation. Examples of such cost functions include the MAP estimate of cardinality variance~\cite{hoang_automatica_2014}, statistical mean of cardinality variance~\cite{amir_ISSNIP_2013}, posterior expected error of cardinality and states~(PEECS)~\cite{amir_SPL_2013,amir_SP_2016,amir_fusion_2014} and statistical mean of the OSPA error~\cite{amir_OSPA_based_TAES_2015}. A general discussion and comparison between task-driven and information-driven objective functions for sensor management is presented in~\cite{comparison_task_inf_2005}. 

In the multi-sensor control framework proposed in this paper, we use PEECS as the objective (cost) function. The rationale behind this choice is that while computing PEECS can be faster than the common divergence functions, comparable or better tracking accuracies can be achieved via minimizing PEECS as the sensor control cost function~\cite{amir_SP_2016,amir_fusion_2014}.

\subsection{Sensor fusion and optimal control}
\label{subsec:fusion}
In presence of multiple sensors (or sensor nodes in a sensor network), usually a multi-object Bayes filter runs at each node and the local posteriors need to be fused. The Generalized Covariance Intersection (GCI) rule has been widely used for consensus-based fusion of multiple multi-object densities of various forms. Examples include the fusion of Poisson multi-object posteriors of multiple local PHD filters~\cite{battistelli_consensus_PHD_SPIE_2015}, i.d.d. clusters densities of several local CPHD filters~\cite{battistelli_consensus_CPHD_2013}, multi-Bernoulli densities of local multi-Bernoulli filters~\cite{bailu_2016}, and LMB or Vo-Vo densities of several local LMB or Vo-Vo filters~\cite{consensus_labelled_RFS_2016}. The problem of multi-sensor control for labeled random set filters is recently considered by Meng~\textit{et~al.}~\cite{meng_CS_2016}. In this method, local Vo-Vo filters are operating at each sensor node, and the resulting Vo-Vo densities (posteriors) are fused using the GCI-rule (as formulated in~\cite{consensus_labelled_RFS_2016}). 

The common underlying assumption for solving the multi-sensor control problem is that in an \textit{exhaustive search} scheme, the objective function is computed for all possible combinations of sensor control actions $\mathfrak{u} = (u_1,\ldots,u_{n_s})$. This approach works well for a relatively small number of sensors. For instance, the case study presented in the work of Meng~\textit{et~al.}~\cite{meng_CS_2016} involves only two sensors. In presence of numerous sensors, their combined control becomes computationally intractable if implemented via an exhaustive search. Indeed, the computational cost of overall multi-sensor control procedure will grow exponentially with the number of sensors. Our framework includes a \textit{guided search} method that solves the optimization problem without the need for an exhaustive search and can be utilized to simultaneously control numerous sensors.

\section{Multi-Sensor Control Framework}
\label{sec:approach}

Given the system model presented in section~\ref{sec:prob_stat}, an effective design for a multi-sensor control framework is presented in this section. We first outline an overview of the general components and steps involved in our proposed approach. Having the big picture in mind, we then present the details of various components as implemented in our experiments. 

Let us assume that at each time $k$, the fused prior from the previous step, $\bm{\tilde{\pi}}_{k-1}$, is processed through the prediction step of the Bayes filter. A multi-object set estimate, $\bm{\hat{X}}_{k|k-1}$ is then extracted from the predicted density and used to compute predicted ideal measurement sets (PIMS)~\cite{amir_SPL_2013,amir_SP_2016,book_Mahler_2014} for each sensor node and each possible control command applied to that node, denoted by $\{Z_i(u)\}_{u\in\mathbb{U}}$ for sensor $i$.

In the next step, at each sensor node, a \textit{pseudo update} is performed using each PIMS associated with a control command. The resulting pseudo posteriors are then processed by an \textit{optimization} module to output an optimal set of control commands. The control actions are then applied to the sensors (for instance, they are displaced or rotated according to the chosen action command) following which, the measurement sets $Z_1, \ldots, Z_{n_s}$ are acquired from the sensors. Using those measurement sets, the predicted multi-object density is locally updated in each sensor node, then the local posteriors are fused using a fusion rule such as the GCI-rule. The fused posterior is post-processed (\textit{e.g.} low weight components are pruned or particles are resampled). The resulting posterior is then used as prior in the next time step.

\subsection{Labeled Multi-Bernoulli filter}
The notion of Labeled Multi-Bernoulli ({LMB}) {RFS} was introduced for the first time in~\cite{vo_glmb_TSP_2013}, with the LMB filter recursion further developed in~\cite{vo_lmb_TSP_2014}. The LMB distribution is completely described by its components ${\pi} = \{(r^{(\ell)},p^{(\ell)}(\cdot))\}_{\ell\in\mathbb{L}}$ where $r^{(\ell)}$ is the \textit{probability of existence} of an object with label $\ell \in \mathbb{L}$, and $p^{(\ell)}(x)$ is the probability density of the object's state $x \in \mathbb{X}$ conditional on its existence. The LMB RFS density is given by
\begin{equation}
\label{eq:LMB_Distribution}
\pi(\mathbf{X})=\Delta (\mathbf{X})w (\mathcal{L}(\mathbf{X}))\left[ p\right] ^{\mathbf{X}},  
\end{equation}
where $\mathcal{L}(\bm{X})$ is the set of all labels extracted from labeled states in $\bm{X}$, and 
\begin{equation}
\Delta(\mathbf{X}) \triangleq
\begin{cases}
1 & \mathrm{if}\ |\mathbf{X}| = |\mathcal{L}(\mathbf{X})| \\
0 & \mathrm{otherwise,}
\end{cases}
\end{equation}
 in which $|\cdot|$ means ``the cardinality of", and 
 \begin{equation}
 [p]^{\bm{X}} \triangleq \prod_{(x,\ell)\in\bm{X}} p^{(\ell)}(x),
 \end{equation}
 and
\begin{eqnarray}
w(L) &=& \prod\limits_{i \in \mathbb{L}}\left( 1-r^{(i)}\right) \prod\limits_{\ell \in L} \frac{1_{\mathbb{L}}(\ell)r^{(\ell)}} {(1-r^{(\ell)})}
\end{eqnarray}
is the probability of joint existence of all objects with labels $\ell \in L$ and non-existence of all other labels~\cite{vo_lmb_TSP_2014}. 

In a Bayes multi-object filter, suppose that the prior is an LMB with parameters $ \{(r^{(\ell)},p^{(\ell)}(\cdot))\}_{\ell\in\mathbb{L}}$. In an SMC implementation, the density function of each component with label $(\ell)$ is approximated by $J^{(\ell)}$ particles and weights, 
\begin{equation}
p^{(\ell)}(x) \approx \sum_{j=1}^{J^{(\ell)}} w_{j}^{(\ell)}
\delta(x-x_{j}^{(\ell)})
\end{equation}
where $\delta(\cdot)$ is the Dirac delta function.

In the prediction step of an LMB filter, the LMB prior is turned into the following new LMB density with evolved particles and probabilities of existence including the LMB birth components:
\begin{equation}
{\pi}_{+} = \left\{\left(r_{+,S}^{(\ell)},p_{+,S}^{(\ell)}\right)\right\}_{\ell\in\mathbb{L}} \cup 
\left\{\left(r_{B}^{(\ell)},p_{B}^{(\ell)}\right)\right\}_{\ell\in\mathbb{B}}
\end{equation} 
where 
\begin{eqnarray}
r_{+,S}^{(\ell)} & = & \eta_S(\ell)\, r^{(\ell)} \\
p_{+,S}^{(\ell)} & = & \langle p_S(\cdot,\ell) f(x|\cdot,\ell),p^{(\ell)}(\cdot) \rangle / \eta_S(\ell)
\end{eqnarray}
and 
\begin{equation}
\eta_S(\ell) = \langle p_S(\cdot,\ell),p^{(\ell)}(\cdot) \rangle.
\end{equation}
Let us denote the predicted LMB parameters by $\{r_+^{(\ell)},\{w_{+j}^{(\ell)},x_{+j}^{(\ell)}\}_{j = 1}^{J_+^{(\ell)}}\}_{\ell\in\mathbb{L}_+}$ where $\mathbb{L}_+ = \mathbb{L}\cup\mathbb{B}$. Note that in above equations, 
$$\langle f , g \rangle  \triangleq \int_{\mathbb{X}} f(x) g(x) dx$$
denotes inner product of two functions.

\begin{Rem}
	As part of the multi-sensor control framework, a multi-object state estimate needs to be computed from the predicted density. A maximum a posteriori (MAP) estimate for the number of objects can be found from cardinality distribution,
	\begin{equation}
	\label{eq:nhat}
	\begin{array}{rcl}
	\hat{n} & = & \underset{n}{\arg\max} \ \ \ \rho(n)\\
	& = & \underset{n}{\arg\max} \ \ \ \rho(0)\!\!  
	\underset{
		L\subseteq \mathbb{L},|L|=n
	}
	{\sum}
	\left(
	\prod_{\ell\in L}
	\frac{r_+^{(\ell)}}{1-r_+^{(\ell)}}\right).
	\end{array}
	\end{equation} 
	where $\rho(0) = \prod_{\ell\in\mathbb{L}} (1-r_+^{(\ell)})$. Given the number of objects, we find the $\hat{n}$ labels with highest probabilities of existence. For each label, 
	an expected a posteriori (EAP) state estimate is given by 
	\begin{equation}
	\label{eq:xhat}
		\hat{x}_{\text{pseudo}}^{(\ell)} = \sum_{j = 1}^{J_+^{(\ell)}} w_{+j}^{(\ell)} x_{+j}^{(\ell)} 
	\end{equation}
	and the set of all estimates is denoted by $\hat{X}_{\text{pseudo}}$. The subscript ``pseudo'' is used because the estimates are resulted from the predicted, and not the updated, density.
\end{Rem}

Assume that at a sensor node $i$, the control command $u \in \mathbb{U}$ is applied, and a measurement set denoted by $Z_i$ is acquired. Let us denote the updated LMB by 
$$
{\pi}_{i,u}(\cdot|Z_i) = \left\{
\left(
r_{i,u}^{(\ell)},p_{i,u}^{(\ell)}(\cdot)
\right)
\right\}_{\ell\in\mathbb{L}_+}.
$$
According to LMB update equations derived in~\cite{vo_lmb_TSP_2014}, the parameters of the above density are given by:
\begin{eqnarray}
r_{i,u}^{(\ell)} & = & \sum_{(I_+,\theta)\in\mathcal{F}(\mathbb{L}_+)\times{\Theta}_{I_+}}
w^{(I_+,\theta)}(Z)\ 1_{I_+}(\ell) \\
p_{i,u}^{(\ell)}(x) & = & \frac{1}{r^{(\ell)}} \sum_{(I_+,\theta)\in\mathcal{F}(\mathbb{L}_+)\times{\Theta}_{I_+}}
\hspace{-3mm}w^{(I_+,\theta)}(Z)\ 1_{I_+}(\ell) p^{(\theta)}(x,\ell) \nonumber
\\
&&
\end{eqnarray}
where
\begin{eqnarray}
w^{(I_+,\theta)}(Z) & \propto & w_{+}(I_+) [\eta_Z^{(\theta)}]^{I_+} \\
p^{(\theta)}(x,\ell) & = & \frac{p_+^{(\ell)}(x) \psi_Z(x,\ell;\theta)}{\eta_Z^{(\theta)}(\ell)} \\
\eta_Z^{(\theta)}(\ell) & = & \langle  p_+^{(\ell)}(x),\psi_Z(x,\ell;\theta) \rangle \\
\psi_Z(x,\ell;\theta) & = & 
\left\{
\begin{array}{lcr}
\frac{p_D(x,\ell) g(z_{\theta(\ell)}|x,\ell)}{\kappa(z_{\theta(\ell)})}, & \mathrm{if} & \theta(\ell)>0 \\
1-p_D(x,\ell), & \mathrm{if} & \theta(\ell) = 0
\end{array}
\right.
\end{eqnarray}
and $\Theta_{I_+}$ is the space of mappings $\theta: I_+ \rightarrow \{0,1,\ldots,|Z|\}$ such that $\theta(i) = \theta(i')>0$ implies $i = i'$,
and the weight term, $w_{+}(I_+)$, is given by:
\begin{equation}
w_{+}(I_+) = \prod_{i\in \mathbb{L}_+} \left(1-r_+^{(i)}\right) 
\prod_{\ell\in I_+}
\frac{1_{\mathbb{L}_+}(\ell) r_+^{(\ell)}}{1-r_+^{(\ell)}}.
\end{equation}

\subsection{Sensor fusion}

During the update step of LMB filter, the particles do not change, and only their weights evolve. Hence, 
	$
	x_{i,u,j}^{(\ell)} = x_{+j}^{\ell}. 
	$
In other words, all the updated LMB posteriors will have the same particles but with different weights and existence probabilities. This makes the fusion of the posteriors generated at each sensor straightforward. 

For sensor fusion purposes, we use the GCI-rule as derived in~\cite{consensus_labelled_RFS_2016,meng_CS_2016} for fusion of multiple LMB densities. For each multi-sensor command candidate $\mathfrak{u} = (u_1,\ldots,u_{n_s}) \in \mathbb{U}^{n_s}$, the corresponding posteriors are LMB's with parameters 
	$
	\left\{
	\{
	(r_{i,u_i}^{(\ell)}, p_{i,u_i}^{(\ell)}(\cdot))
	\}_{\ell\in\mathbb{L}}
	\right\}_{i=1}^{n_s}
	$
where each density is approximated by the same particles but different weights,
	\begin{equation}
	p_{i,u_i}^{(\ell)}(x) \approx \sum_{j=1}^{J_+^{(\ell)}} w_{i,u_i,j}^{(\ell)}\delta(x-x_{+,j}^{(\ell)}).
	\label{eq:den_approx}
	\end{equation}

The GCI-rule returns the following fused existence probabilities and densities:
\begin{flalign}
r_{\mathfrak{u}}^{(\ell)} = & \frac{\int \prod_{i=1}^{n_s} \left(r_{i,u_i}^{(\ell)} p_{i,u_i}^{(\ell)}(x)\right)^{\omega_i}dx}{
	\prod_{i=1}^{n_s}\left(1-r_{i,u_i}^{(\ell)}\right)^{\omega_i}
	+\int \prod_{i=1}^{n_s} \left(r_{i,u_i}^{(\ell)} p_{i,u_i}^{(\ell)}(x)\right)^{\omega_i}dx} 
\label{eq:fused_r_1}
\\
p_{\mathfrak{u}}^{(\ell)}(x) = & \frac{
	\prod_{i=1}^{n_s} \left(p_{i,u_i}^{(\ell)}(x)\right)^{\omega_i}
}
{\int \prod_{i=1}^{n_s} \left(p_{i,u_i}^{(\ell)}(x)\right)^{\omega_i} dx}
\label{eq:fused_p_1}
\end{flalign}
where $\omega_i$ is a constant weight indicating the strength of our emphasis on sensor $s_i$ in the fusion process. These weights should be normalized, i.e. $\sum_{i=1}^{n_s} \omega_i = 1$. In our simulation studies, we assumed that all sensor nodes have equal priority, and used the values $\omega_i = \frac{1}{n_s}$.

Substituting each density with its particle approximation~\eqref{eq:den_approx} turns the integrals to weighted sums over the particles. It is here that sharing the same particles between all the densities becomes instrumental for computation of fused parameters. The fused existence probability is given by:
\begin{equation}
r_{\mathfrak{u}  }^{(\ell)} = \frac{
	\sum_{j=1}^{J_+^{(\ell)}} \prod_{i=1}^{n_s} \left(r_{i,u_i}^{(\ell)} w_{i,u_i}^{(\ell)}\right)^{\omega_i}}
{
	\prod_{i=1}^{n_s}\left(1-r_{i,u_i}^{(\ell)}\right)^{\omega_i}
	+\sum_{j=1}^{J_+^{(\ell)}} \prod_{i=1}^{n_s} \left(r_{i,u_i}^{(\ell)} w_{i,u_i}^{(\ell)}\right)^{\omega_i}}.
\label{eq:fused_r_2}
\end{equation}
The fused densities also take the form of weighted sum of Dirac deltas:
\begin{equation}
p_{\mathfrak{u}  }^{(\ell)}(x) = \sum_{j=1}^{J_+^{(\ell)}} w_{\mathfrak{u}  ,j}^{(\ell)}\ \delta(x-x_{+j}^{(\ell)})
\label{eq:fused_p_2}
\end{equation}
where 
\begin{equation}
w_{\mathfrak{u}  ,j}^{(\ell)} = 
\frac{
	\prod_{i=1}^{n_s} \left(w_{i,u_i,j}^{(\ell)}\right)^{\omega_i}
}
{ 
	\sum_{j=1}^{J_+^{(\ell)}}
	\prod_{i=1}^{n_s} \left(w_{i,u_i,j}^{(\ell)}\right)^{\omega_i}
}
\label{eq:weight_GCI_fusion}
\end{equation}
is the fused weight of each particle in the fused pseudo-posterior. 

\begin{Rem}
	To maintain tractability, LMB components with extremely small existence probabilities should be pruned. This is performed after GCI-fusion of the posteriors.
\end{Rem}

\subsection{Objective function}
For the objective function, we chose the task-driven cost function termed PEECS in~\cite{amir_SP_2016}. It returns a linear combination of the cardinality and state estimation errors which are quantified by computing the variance of cardinality and weighted sum of single-object variances, respectively. Consider the fused LMB posterior parametrized by
$
\pi_{\mathfrak{u}  } = \left\{ (r_{\mathfrak{u}  }^{(\ell)},p_{\mathfrak{u}  }^{(\ell)}(\cdot)
)
\right\}_{\ell\in \mathbb{L}_+}
$   
where
$
p_{\mathfrak{u}  }^{(\ell)}(x) = 
\sum_{j=1}^{J_+^{(\ell)}} w_{\mathfrak{u}  ,j}^{(\ell)}\ \delta(x-x_{+j}^{(\ell)}).
$
The PEECS cost associated with the multi-sensor control choice $\mathfrak{u}  $ is then given by:
\begin{equation}
\nu(\mathfrak{u}  ;\pi_{\mathfrak{u}}) = \eta \epsilon^2_{|X|} + (1-\eta) \epsilon^2_X
\end{equation} 
where $\eta$ is a user-defined parameter representing the level of emphasis on desired accuracy of number of targets versus accuracy of state estimates, and
\begin{equation}
\epsilon^2_{|X|} = \sum_{\ell\in \mathbb{L}_+} r_{\mathfrak{u}  }^{(\ell)} (1-r_{\mathfrak{u}  }^{(\ell)})
\textrm{;}\ \ \ 
\epsilon^2_{X} = \frac{\sum_{\ell\in \mathbb{L}_+} r_{\mathfrak{u}  }^{(\ell)} \sigma_{X^{(\ell)}}^2}
{\sum_{\ell\in \mathbb{L}_+} r_{\mathfrak{u}  }^{(\ell)}}
\end{equation}
in which we have:
\begin{equation}
\sigma_{X^{(\ell)}}^2 = 
\sum_{j=1}^{J_+^{(\ell)}} w_{\mathfrak{u}  ,j}^{(\ell)} \left(x_{+j}^{(\ell)}\right)^2 -
\left( \sum_{j=1}^{J_+^{(\ell)}} w_{\mathfrak{u}  ,j}^{(\ell)} x_{+j}^{(\ell)}\right)^2.
\end{equation}

\subsection{Optimization for multi-sensor control}
The final step to solve the control problem in POMDP framework is to find the optimum point of the objective function. The common approach, which is usually tractable with few sensors, is based on an exhaustive grid search in the discrete multi-sensor control command space. In this approach, for all possible $n_s$-tuples 
$$
\mathfrak{u} = (u_1,\ldots,u_{n_s}) \in \mathbb{U}^{n_s}
$$
an ideal measurement set (PIMS)~\cite{Mahler_sensors} is synthetically generated from the prediction at each sensor node, and using the PIMS, a local LMB update is run to create a  pseudo-posterior. For each possible multi-sensor control command $\mathfrak{u}$, the corresponding pseudo posteriors are fused and the objective function is computed. The optimal sensor control decision is then given by:
\begin{equation}
(u^*_1,\ldots,u^*_{n_s}) = \underset{{(u_1,\ldots,u_{n_s}) \in \mathbb{U}^{n_s}}}{\arg\min} \nu(u_1,\ldots,u_{n_s};\pi_{k|k-1})
\end{equation}
where $\nu(u_1,\ldots,u_{n_s};\pi_{k|k-1})$ denotes the objective function computed from the fused pseudo-posterior via updating the predicted density $\pi_{k|k-1}$ if control actions $(u_1,\ldots,u_{n_s})$ are applied. Note that in the above equation (and the rest of this paper), the objective function is assumed to be a cost. If it is a reward, its optimization would require maximization.

The above search requires the ``fusion of pseudo-posteriors followed by computation of the objective function'' to be repeated for $|\mathbb{U}|^{n_s}$ times where $|\mathbb{U}|$ denotes the cardinality of single sensor control commands space $\mathbb{U}$. The computational cost increases exponentially with the number of sensors, and becomes intractable when a relatively large number of sensors are involved. 

We propose a guided search routine inspired by the coordinate descent method that significantly accelerates the optimization process and makes it suitable for real time implementation. Our guided search is an iterative coordinate descent type method with random initializations. Coordinate descent algorithms are well-known for their simplicity, computational efficiency and scalability. An overview of coordinate descent algorithms for various optimization problems with different constraints is presented in~\cite{coordinate_descent_Wright_2015}. These algorithms are derivative-free and perform a line search along one coordinate direction at the current point in each iteration and use different coordinate directions cyclically to find a local optimum point.

Coordinate descent provides a sub-optimal solution with non-differentiable objective functions. \cite{coordinate_descent_Tseng_2001} considers convergence of coordinate descent methods to a stationary, but not necessarily minimum, point for objective functions that include a non-differentiable and separate contribution (also called ``non-smooth part"). In a later work, Spall~\cite{Spall2012CyclicSP} analyzes the convergence of more general seasaw processes for optimization and identification, showing that under reasonable conditions, the cyclic scheme converges to the optimal joint value for the full vector of unknown parameters (sensor commands, in the context of our work).

To find the best $n_s$-tuple of control commands $\mathfrak{u}   = (u_1,\ldots,u_{n_s})$ in the $n_s$-dimensional command space $\mathbb{U}^{n_s}$, our guided search starts with random initialization of control commands, denoted by $\mathfrak{u}  ^0 = (u_1^0,\ldots,u_{n_s}^0)$. We then solve the optimization problem
\begin{equation}
u_1^1 = \underset{u}{\arg\min}~\nu(u, u_2^0,\ldots,u_{n_s}^0;\pi_{k|k-1})
\end{equation}
via exhaustive search in the space of coordinates associated with sensor 1. We replace the candidate multi-sensor control action with $(u_1^1,u_2^0,\ldots,u_{n_s}^0)$. Repeating the one-dimensional search for all the other coordinates associated with different sensors, our candidate turns into $\mathfrak{u}  ^1 = (u_1^1,\ldots,u_{n_s}^1)$. We repeat this cycle over to obtain the next candidates $\mathfrak{u}  ^2, \mathfrak{u}  ^3, \ldots$ until convergence, i.e. until we find $n$ for which $\mathfrak{u}  ^{n-1} = \mathfrak{u}  ^n$. When used in such an iterative (cyclic) routine, the search is proven to converge in finite time~\cite{coordinate_descent_Saha_2010}. 

The converged $n_s$-tuple can be a local optimum. Hence, we need to repeat the process with multiple random initializations and choose the best candidate as the multi-sensor control command, $\mathfrak{u}  ^*$.  
The required number of repeated convergence with random initializations depends on the number of local optima and the desired chance of success. If there are $\mathfrak{M}$ local optima, in the worst case scenario they all have basins of attraction with the same hyper-volume, i.e. each basin of attraction is comprised of $\frac{1}{\mathfrak{M}}$ of all the points in $\mathbb{U}^{n_s}$. Thus, the chance of a randomly initialized search converging to the global optimum will be $\frac{1}{\mathfrak{M}}$.\footnote{Usually, the basin of attraction for the global optimum is larger and the chance of success in each round of random initialization is greater than $\frac{1}{\mathfrak{M}}$.} With $N$ random initializations, the total chance of success is $P_{\mathrm{success}} = 1-(1-\frac{1}{\mathfrak{M}})^{N}$. Hence, the required number of random initializations is given by
\begin{equation}
N = 
\left\lceil
\frac{\log(1-P_{\mathrm{success}})}{\log(1-\frac{1}{\mathfrak{M}})}
\right\rceil
\label{eq:no_rand_init}
\end{equation}   
where $\lceil\cdot\rceil$ means rounding up to the next integer. 

In our experiments, choosing the number of local optima at $\mathfrak{M} = 2 n_s$ led to sufficient random initializations for satisfactory results. Based on equation~\eqref{eq:no_rand_init}, with a probability of success of 95\%, for $n_s = $5, 10 and 20 sensors, we would require $N =$~29,~59~and~119 random initializations which need far less computation than exhaustive search in the multi-dimensional space $\mathbb{U}^{n_s}$. Interestingly, the required number of initializations in equation~\eqref{eq:no_rand_init} does not depend on $|\mathbb{U}|$, i.e. it does not increase with the resolution of the sensor command space.

\subsection{Step-by-step Algorithm}
Algorithm~\ref{alg:1} shows a complete step-by-step pseudocode for multi-sensor control within the LMB filter, that outputs a fused posterior. Starting with an LMB prior (which is the fused LMB posterior from previous time), the function $\textsc{Predict}(\cdot)$ implements the LMB prediction step. Multiple object states are then estimated from the predicted LMB density by calling the function $\textsc{Estimate}(\cdot)$ which implements equations~\eqref{eq:nhat}~and~\eqref{eq:xhat}.

The coordinate descent guided search for multi-sensor control is implemented through the line numbers 3--24 in Algorithm~\ref{alg:1}. Before the search begins, for every sensor, $s_i$ and every possible action command $u$, a PIMS is computed. Using that set of ideal measurements, the LMB density is then updated by calling function $\textsc{Update}(\cdot)$, and its parameters (existence probabilities $r_{i,u}^{(\ell)}$, particles $x_{i,u,j}^{(\ell)}$ and their weights $w_{i,u,j}^{(\ell)}$) are recorded. 

The function $\textsc{Cost}$ computes the PEECS cost value for each set of local posteriors associated with multiple sensor control commands. Both within the cost computation steps, and at the conclusion of the Algorithm~\ref{alg:1}, we need to apply the GCI-rule to fuse the locally (pseudo-)updated LMB posteriors. 
The function $\textsc{GCIfusion}$ performs this task.

\begin{algorithm*}
	\caption{Step-by-step pseudocode for a single iteration of filtering, fusion and multi-sensor control. \label{alg:1}}
	\includegraphics[width=6.4 in]{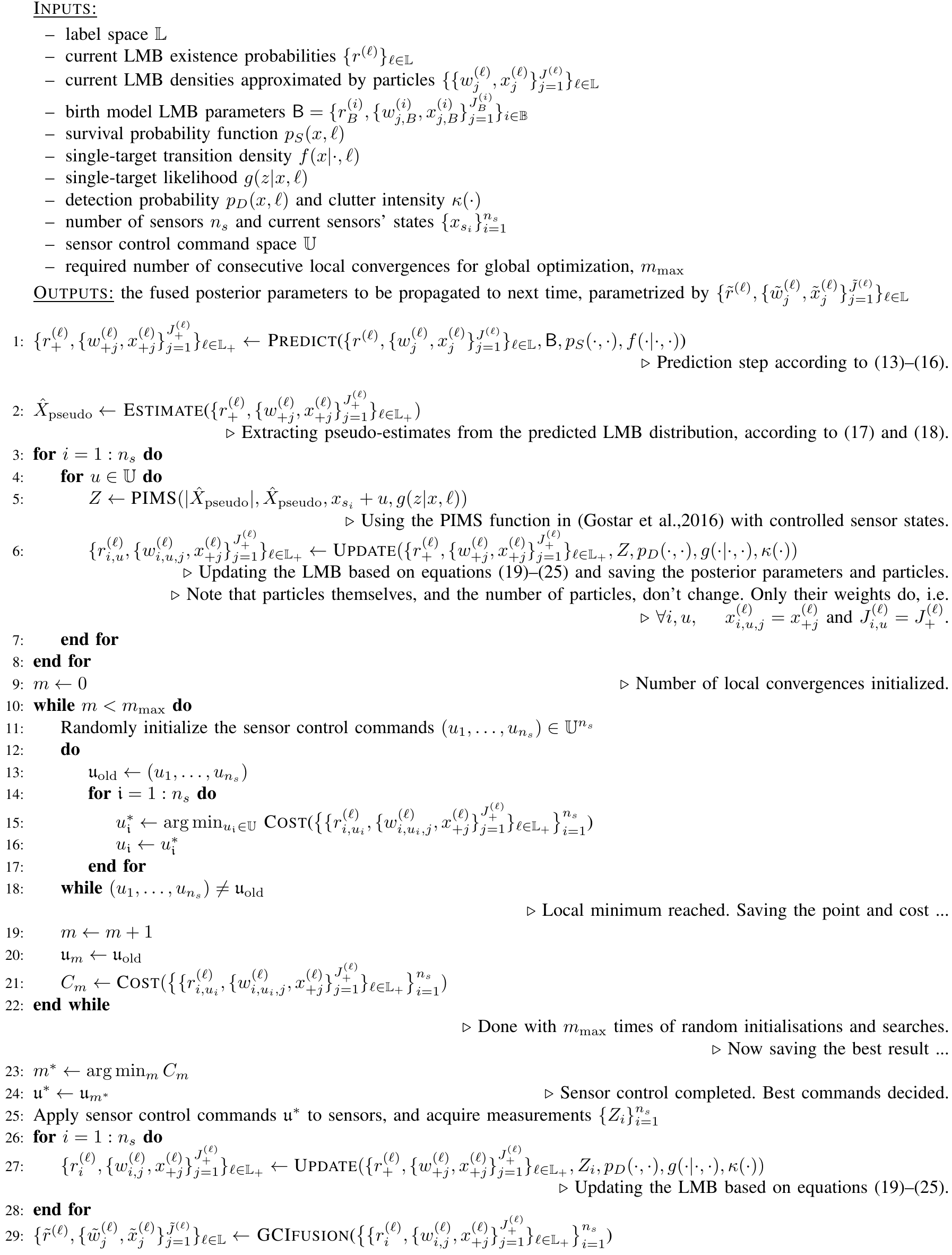}
\end{algorithm*}

\section{Simulation Results}
\label{sec:sim_res}

We conducted an extensive set of experiments involving various scenarios with different numbers of targets, sensors, target motion models and sensor detection profile models. In each experiment, we compared the performance of the proposed  multi-sensor control solution with the exhaustive search-based method (similar to~\cite{meng_CS_2016}), in terms of both accuracy and computational cost. This section includes representative set of our simulation results. Those show the advantages of the proposed method, particularly, in terms of computation time. All scenarios share the following parameters.

The targets maneuver in an area of $1200~\textrm{m}~\times~1200~\textrm{m}$. The single target state $\bm{x}$ is comprised of its label and unlabeled state. The label is formed as $\ell = (k_B,i_B)$ where $k_B$ is the birth time of the target and $i_B$ is an index to distinguish targets born at the same time. The unlabeled state is four-dimensional and includes the Cartesian coordinates of the target and its speed in those directions, denoted by $x = [p_x\ \dot{p}_x\ p_y\ \dot{p}_y]^\top$. Each target moves according to the Nearly-Constant Velocity (NCV) model with its variance parameter denoted by  $\sigma_w^2$. With this model, the transition density is $f_{k|k-1}(x_k|x_{k-1},\ell)=\mathcal{N}(x_k;F_kx_{k-1},Q_k)$, where 
\begin{equation}
F_k=
\begin{bmatrix}
1 & T & 0 & 0\\
0 & 1 & 0 & 0\\
0 & 0 & 1 & T\\
0 & 0 & 0 & 1
\end{bmatrix},
Q_k=\sigma_w^2
\begin{bmatrix}
\frac{T^3}{3} &  \frac{T^2}{2}   & 0 & 0\\
\frac{T^2}{2}  & T& 0 & 0 \\
0  & 0 & \frac{T^3}{3}  & \frac{T^2}{2}\\
0 & 0  &\frac{T^2}{2} & T
\end{bmatrix}
\end{equation}
and $T$ is the sampling interval~(in our experiments $T=1\,$s). The probability of survival is fixed at $p_S(x,\ell) = 0.99$. For each sensor $s_i$, with its location denoted by $[x_{s_i}\ y_{s_i}]^\top$, each target (if detected) leads to a bearing and range measurement vector with the measurement noise density given by $ \mathcal{N}(\cdot; [0\ 0]^\top, R)$ in which $R=$~diag$(\sigma_\theta^2,\sigma_r^2)$ with $\sigma_\theta^2=2\pi/180$~rad and $\sigma_r^2=10$~m$^2$ being the scales of range and bearing noise. Thus, the single target likelihood function is ~$g(z_i|x,\ell)=\mathcal{N}(z;\mu_i(x),R)$, where 
\begin{equation}
\mu_i(x) = \left[\arctan\left({\frac{p_x\!-\!x_{s_i}}{p_y\!-\!y_{s_i}}}\right)\ \sqrt{({p_x\!-\!x_{s_i}})^2+({p_y\!-\!y_{s_i}})^2}\right]^\top.
\end{equation}

Each measurement set acquired from each sensor also includes Poisson distributed clutter with the fixed clutter rate of $\lambda_c = 5.$ In all scenarios, the density $p^{(\ell)}(\cdot)$ of each labeled Bernoulli component in the filter is approximated by $J^{(\ell)} = 1000$ particles. All simulation experiments were coded using MATLAB R2015b and ran on an Intel® Core™ i7-4770 CPU @3.40GHz, and 8~GB memory. 

\subsection*{Scenario 1: Pseudo-stationary targets}

In this scenario, we tried the commonly used case study in which five targets move with relatively small displacements (are pseudo-stationary). To realize such movements, we applied the NCV motion model with the very small variance $\sigma_w^2=5\times 10^{-2}$~m$^2/$s$^3$ borrowed from similar simulations reported in~\cite{ristic_reward_PHD_TAES_2011,amir_SPL_2013,amir_SP_2016}. In this scenario, the detection profile of each sensor is range-dependent. The detection probability of target with the state $x$ by sensor $s_i$ is given by:
\begin{equation}
p_D^i(x,\ell)= 
\begin{cases}
1 & \mathrm{if}\ d_i(x)\leqslant R_0\\
1-\frac{d_i(x)-R_0}{\eta} & \mathrm{if}\ R_0<d_i(x)\leqslant R_0+\eta\\
0 & \mathrm{otherwise}
\end{cases}
\end{equation}
where $R_0=200$~m, and $\eta=1000$~m denotes the maximum range of detection, and $d_i(x)$ denote the sensor-target distance given by:
\begin{equation}
d_i(x) = \sqrt{(p_x-x_{s_i})^2+(p_y-y_{s_i})^2}.
\end{equation}
The detection probability decreases with increasing sensor-target distance. Because of this, and considering that the targets stay almost in the same distance away from each other all the time, the sensor control is intuitively expected to drive all the sensors towards the center of the pseudo-stationary targets.

The birth process is modeled  by an LMB density with $|\mathbb{B}| = 5$ components. Each component has the same existence probability of $r_B^{(i)}=0.05$, and a Gaussian density $p_B^{(i)}=\mathcal{N}(x;m_B^{(i)},P_B)$, where the mean and covariance of Gaussians are 
$$
\begin{array}{cccccc}
	m_B^{(1)} &=& [800\ 0\ 600\ 0]^\top; &
	m_B^{(2)} &=& [650\ 0 \ 500\ 0]^\top; \\
	m_B^{(3)} &=& [620\ 0\ 700\ 0]^\top; &
	m_B^{(4)} &=& [750\ 0\ 800\ 0]^\top; \\
	m_B^{(5)} &=& [700\ 0\ 400\ 0]^\top; &&& \\
	P_B &=& \multicolumn{4}{l}{\text{diag}\big(1, 5\times 10^{-5}, 1, 5\times\!10^{-5}\big).}
\end{array}
$$

\begin{figure}
\centering
\includegraphics[width=2.2 in]{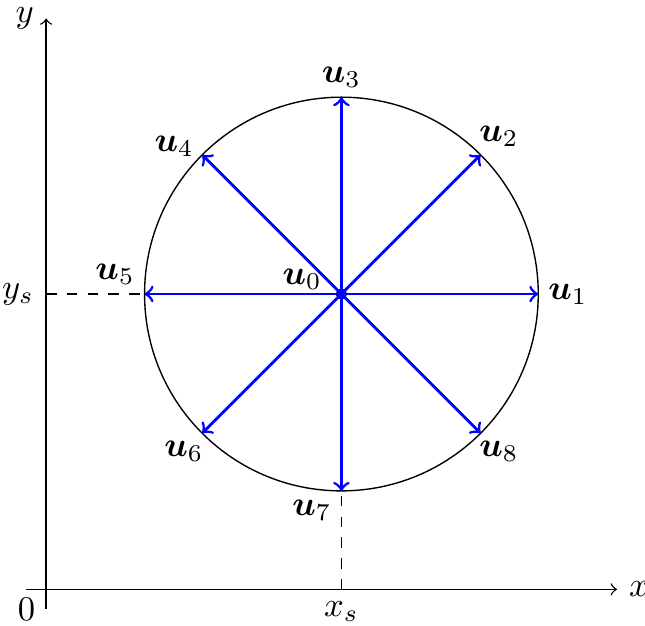}
	\caption{\label{fig:sensor_disps} Nine possible sensor displacements. Note that $\mathbb{U} = \{\mathfrak{u}  _{\mathfrak{j}}\}_{\mathfrak{j}=0:8}$ where $\mathfrak{u}  _0$ denotes zero displacement.}
\end{figure}

Each sensor $s_i$ can be displaced by the multi-sensor control to one of the following possible displacement commands $u\in\mathbb{U}$:
$$
\mathbb{U} = 
\left\{
\begin{bmatrix}
0 \\ 0
\end{bmatrix}
\right\}
\cup
\left\{
\begin{bmatrix}
\Delta  R\cos(\mathfrak{j}\,\Delta\theta) \\ 
\Delta R\sin(\mathfrak{j}\,\Delta\theta)
\end{bmatrix}
\right\}_{\mathfrak{j}=0,...,N_\theta-1}
$$
where $\Delta_R=50$\,m , $N_\theta=8$ and $\Delta_\theta=2\pi/N_\theta$. Thus, nine control actions are possible at each time step as shown in~Fig.~\ref{fig:sensor_disps}.

\begin{figure}
\centering
\begin{tabular}{c}
\includegraphics[width=3 in]{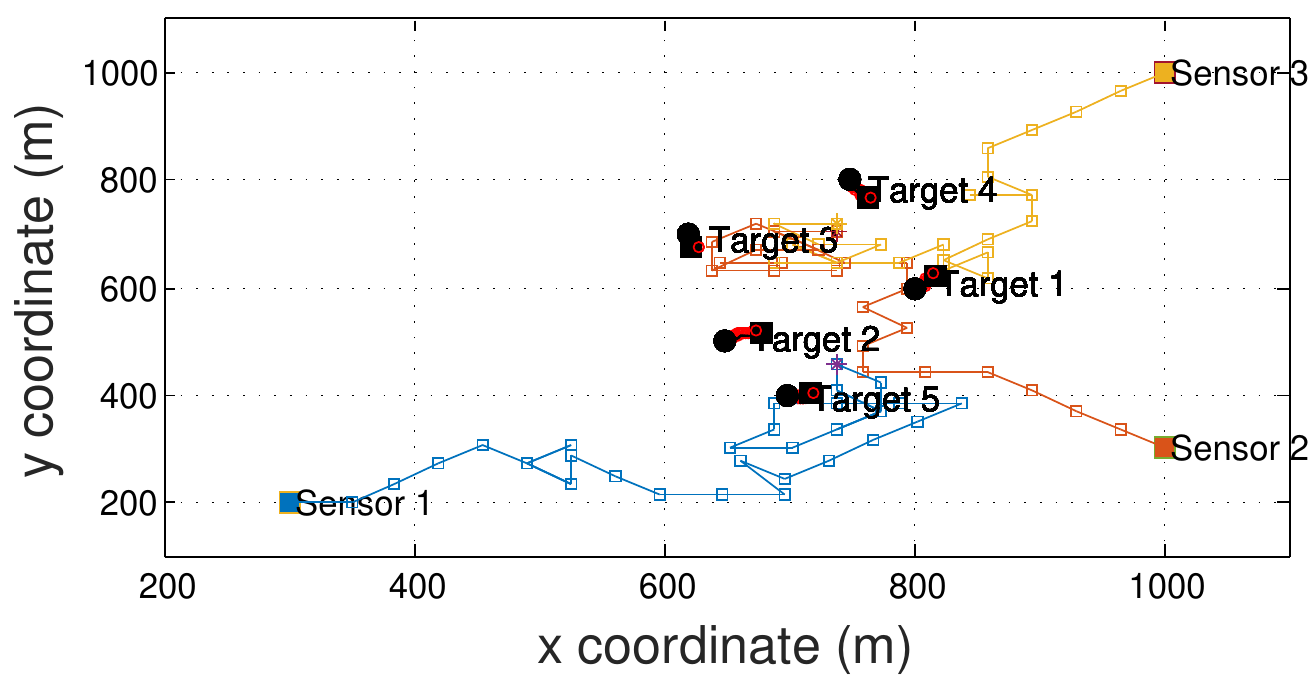}\\
\footnotesize(a)\\
\includegraphics[width=3 in]{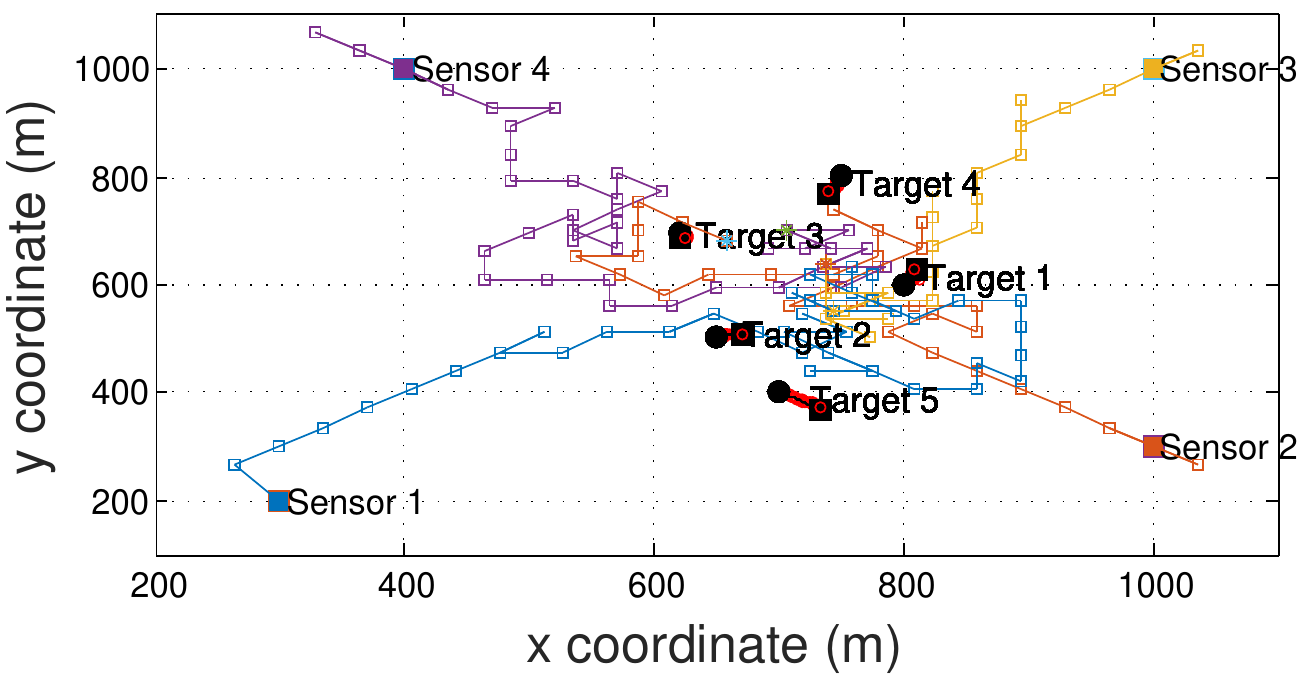}\\
\footnotesize(b)
\end{tabular}
\caption{\label{fig:Sensor_movement} Sensor movements in scenario 1 for (a) three sensors and (b) four sensors. The start and end points for each sensor movement are denoted by colored $\blacksquare$ and $\ast$, respectively. These figures are best viewed in color.}
\end{figure}

Figures~\ref{fig:Sensor_movement}~(a) and~\ref{fig:Sensor_movement}~(b) show the sensor movements in cases with three and four sensors, respectively. As expected, our proposed multi-sensor control method drives all the sensors towards the center of the five targets. 

\begin{figure}
\centering
\begin{tabular}{c}
\includegraphics[width=3 in]{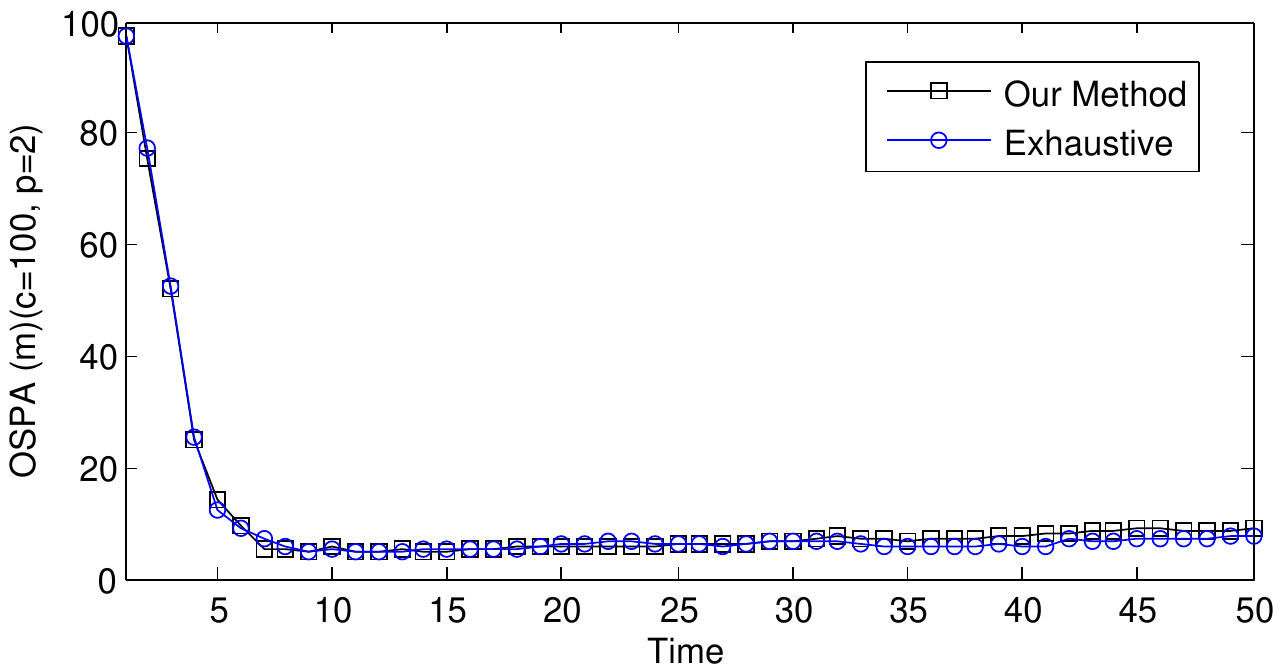}\\
\footnotesize(a)\\
\includegraphics[width=3 in]{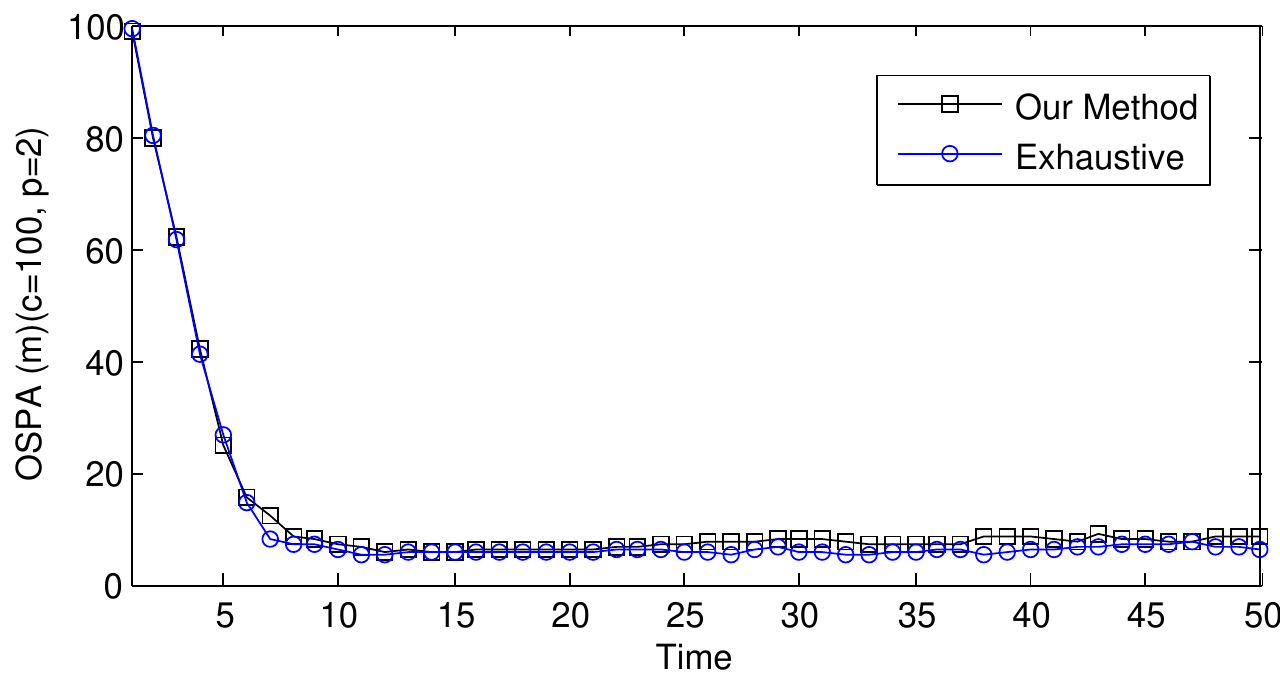}\\
\footnotesize(b)\\
\includegraphics[width=3 in]{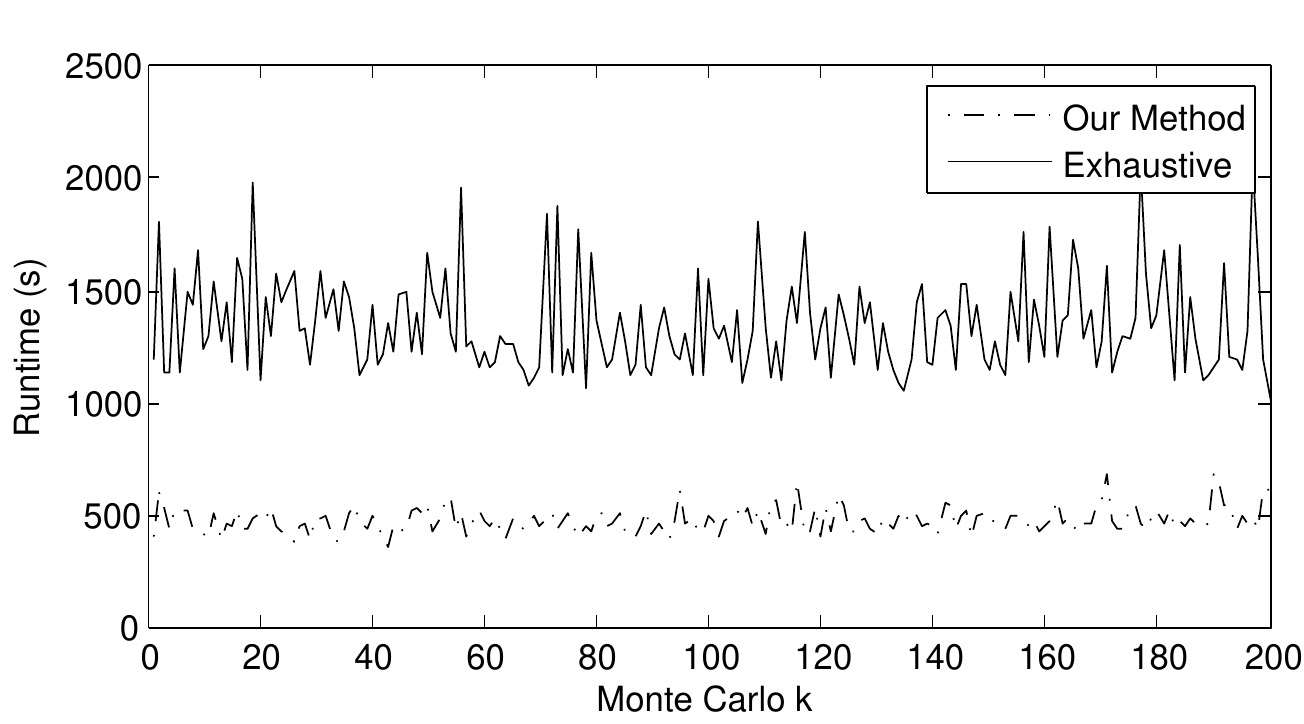}\\
\footnotesize(c)\\
\includegraphics[width=3 in]{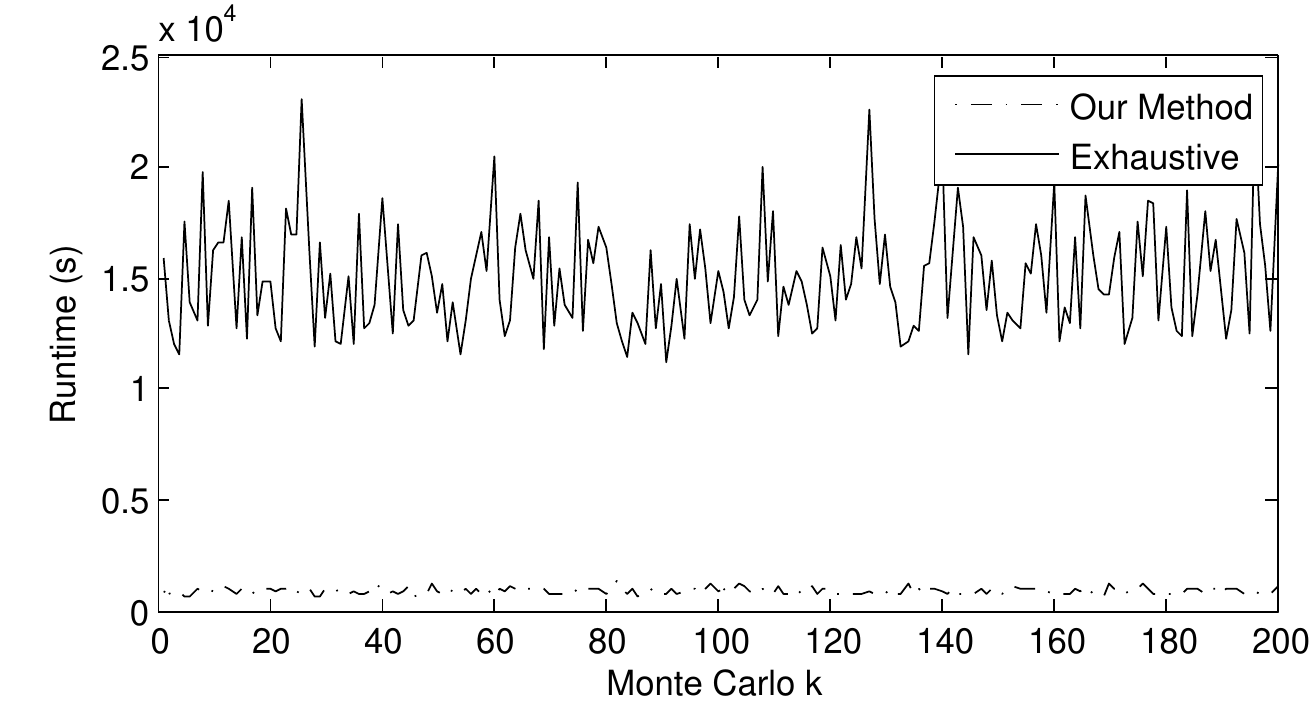}\\
\footnotesize(d)
\end{tabular}
\caption{\label{fig:Sensor_ospas} (a) Average OSPA errors in scenario 1 for three sensors and (b) four sensors. (c) Run times for each time-step (averaged over 200 Monte Carlo runs) for three sensors and (d) four sensors.}
\end{figure}

To quantify and compare the performance of our method, we ran 200 Monte Carlo repetitions, and computed the average OSPA errors with order and cut-off parameters $p=2$ and $c=100$ (see~\cite[Eqs.~(3)-(4)]{vo_OSPA_TSP_2008} for definition of OSPA and its parameters). We also ran the same Monte Carlo experiments but with exhaustive search-based sensor control (same approach in the state of art~\cite{meng_CS_2016}). Figures~\ref{fig:Sensor_ospas}(a) and~\ref{fig:Sensor_ospas}(b) demonstrate that in terms of multi-target tracking errors, the proposed method is comparable with the state of art. The run times for each time-step (averaged over 200 Monte Carlo runs and shown in Figs.~\ref{fig:Sensor_ospas}(c) and ~\ref{fig:Sensor_ospas}(d), however, demonstrate that our method runs three times faster than the exhaustive search method used in~\cite{meng_CS_2016} in presence of three sensors, and 17 times faster when there are four sensors.
 
\begin{figure}
\centering
\includegraphics[width=3 in]{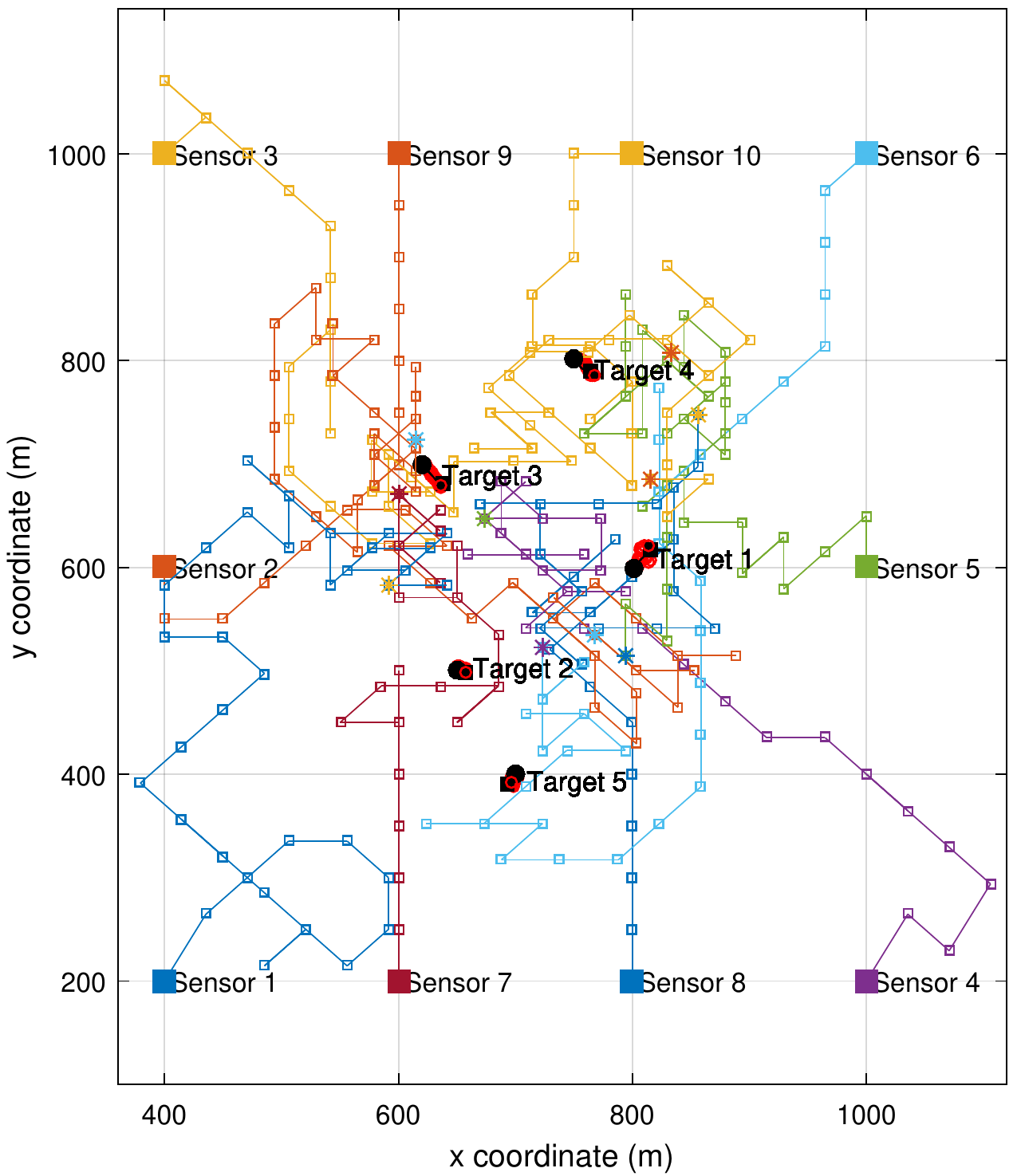}
\caption{\label{fig:10sensor} Controlled movements of 10 sensors in scenario 1: The sensors generally approach the center of targets as expected. Best viewed in color.}
\end{figure}

We also tried the experiment with 10 sensors but the exhaustive search-based method turned out to be intractable in our system (only up to five sensors are feasible). Our accelerated solution however, succeeded with sensors moving generally towards the center of targets as expected (See Fig.~\ref{fig:10sensor}), and the OSPA errors were reasonably small---similar to the results shown in Figs.~\ref{fig:Sensor_ospas}(a) and ~\ref{fig:Sensor_ospas}(b). 

\begin{figure}
\centering
\includegraphics[width=3 in]{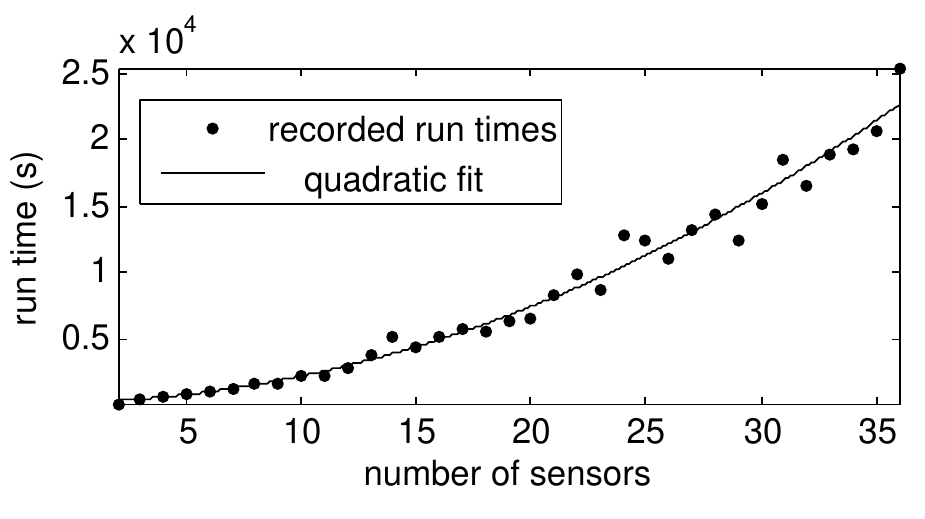}
\caption{\label{fig:runtimes} Recorded run times for scenario 1 in presence of various numbers of sensors.}
\end{figure}

We also applied our method to control various numbers of sensors in the same multi-target tracking scenario. We tried up to 36 sensors, and for each case, recorded the run times as plotted in Fig.~\ref{fig:runtimes}. The results show that with increasing the number of sensors, the run time increases almost quadratically (the best quadratic fit is also displayed). Indeed the computational complexity of our method is almost $\mathcal{O}(n_s^2)$, which is significantly lower than exhaustive search-based multi-sensor control, i.e.~ $\mathcal{O}(|\mathbb{U}|^{n_s})$.

\subsection*{Scenario 2: Maneuvering targets}
In this section, we present the results of multi-sensor control for targets that maneuver with a relatively high speed. In such cases, the sensors are expected to follow and possibly approach the center of the moving targets. We present the results of six sensors controlled to track five targets. For the purpose of visualization of the sensor control performance, we tuned the motion model parameters of the targets in such a way that they move  approximately in the same direction with the same speed. To achieve such target maneuvers, we used the NCV motion model but with the following covariance matrix:
\begin{equation}
Q_k = 
\begin{bmatrix}
B & \bm{0}_2 \\
\bm{0}_2 & B
\end{bmatrix}\ \ ;\ \ B = 
\begin{bmatrix}
0.1 & 0.001 \\ 0.1 & 0.001
\end{bmatrix}.
\end{equation}
Also the birth model parameters were different from scenario~1, as listed below:
$$
\begin{array}{cccccc}
	m_B^{(1)} &=& [1100\ 0\ 200\ 0]^\top; &
	m_B^{(2)} &=& [1200\ 0 \ 300\ 0]^\top; \\
	m_B^{(3)} &=& [1100\ 0\ 300\ 0]^\top; &
	m_B^{(4)} &=& [1200\ 0\ 400\ 0]^\top; \\
	m_B^{(5)} &=& [1200\ 0\ 200\ 0]^\top; &&& \\
	P_B &=& \multicolumn{4}{l}{50\,I_4}
\end{array}
$$
where $I_4$ denotes the four-dimensional identity matrix.

\begin{figure}
\centering
	\includegraphics[width=3 in]{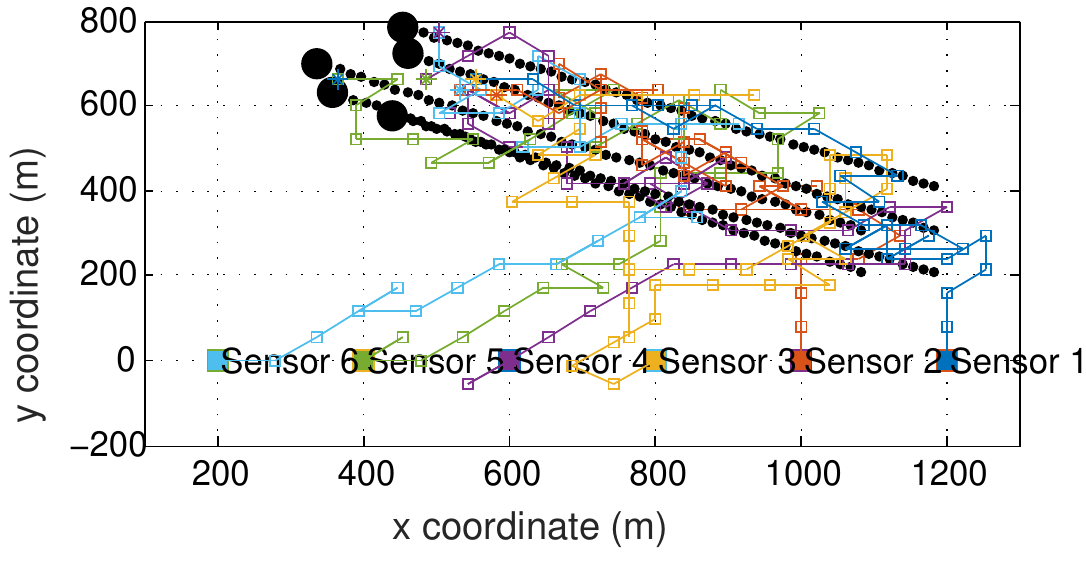}
	\caption{\label{fig:disp_snap} Target and controlled sensor paths in scenario 2 for displacement sensor control actions.}
\end{figure}

We examined the performance of our proposed multi-sensor control method, first with six sensors in a similar fashion to scenario 1 with parameters of state-dependent detection probability to be
$
R_0 = 320\,\mathrm{m}; \eta = 4000\,\mathrm{m}.
$
A snapshot of the final target locations and their paths as well as the controlled sensors and their paths are shown in Fig.~\ref{fig:disp_snap}. It clearly shows how the sensors move and converge to follow the targets. A video of the simulation is available as supplementary material.

\subsection*{Scenario 3: Maneuvering targets and spinning sensors}

\begin{figure}
	\centering
	\includegraphics[width=1.5 in]{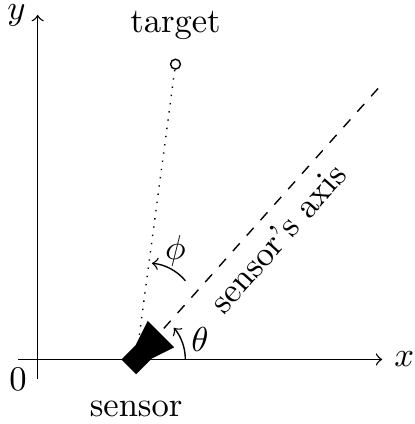}
	\caption{\label{fig:angle_control_pd} Schematics of notations used to formulate control action space and detection probability in scenario 3 with spinning control actions.}
\end{figure}

The proposed multi-sensor control method is not limited to displacement sensor control actions only. The control actions can have other forms. For instance, the sensors can be spun to control angles. We ran a simulation with six sensors that could spin in the interval of 0$^\circ$ to +180$^\circ$. With these sensors, the detection profile is angle-related. For each sensor, the control action command is an axis angle to which the sensor would spin when the control action is applied. Denoting the angle of direction by $\theta$, we considered the action command space of 
$\mathbb{U} = \left\{ \left(\frac{j}{16}\right)\times 180^\circ\right\}_{j=0:16}.
$
The detection probability was assumed to vary with the relative angle of the target with respect to the sensor's axis direction, denoted by $\phi$ as shown in Fig.~\ref{fig:angle_control_pd}. The variations were modeled  as follows:
\begin{equation}
p_D(\phi) = 99.95\% \times \left(1-\frac{\mod{(|\phi|,180^\circ)}}{2000^\circ}\right).
\end{equation}

\begin{figure}
	\centering
	\includegraphics[width=3 in]{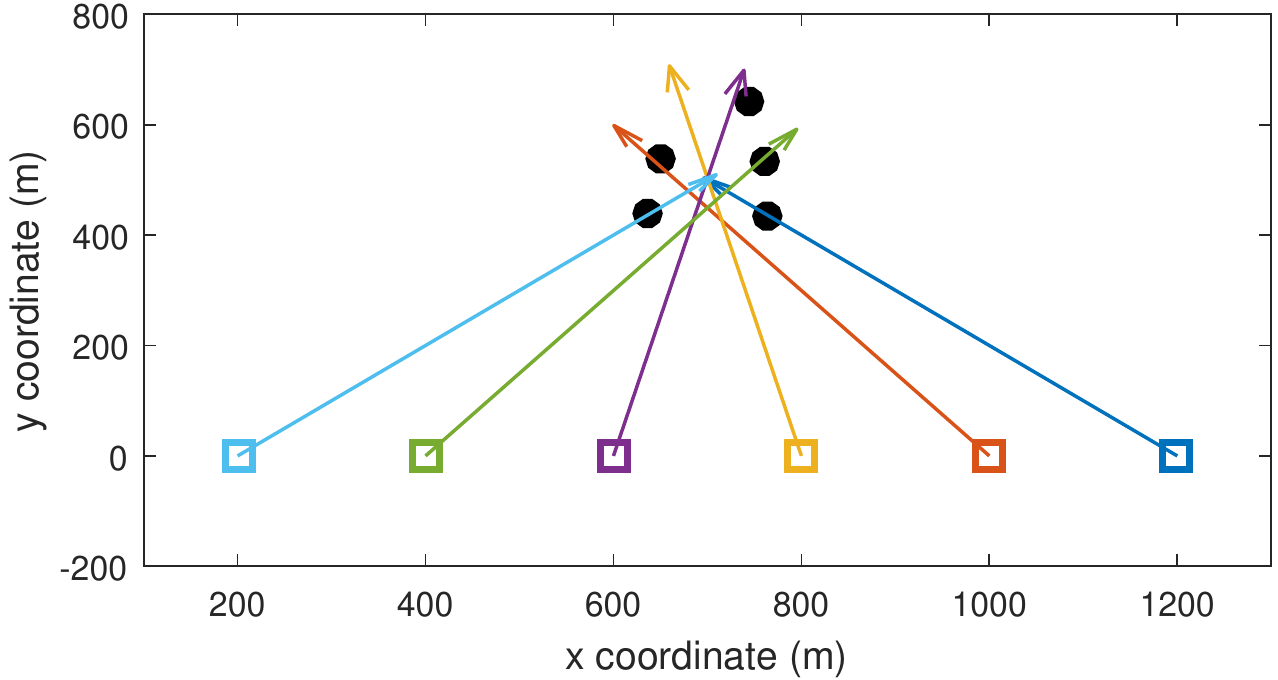}
	\caption{\label{fig:rot_snap} A snapshot of the maneuvering target locations and the controlled angles of sensors in scenario 3.}
\end{figure}

Figure~\ref{fig:rot_snap} shows a snapshot of the targets and how the sensors' axes have been controlled to point towards them. 
A video of the simulation is available as supplementary material that demonstrates the continuous spinning of the sensors in such a way that in general they all point towards the group of targets moving in the scene.

\section{Conclusions}
\label{sec:conc}

A complete POMDP framework for devising multi-sensor control solutions in stochastic multi-object systems was introduced, and a suitable set of choices for various components of the proposed POMDP were outlined. Details of one possible implementation were presented in which the multi-object state is modeled as an LMB RFS, and the SMC implementation of the LMB filter is employed. The proposed framework makes use of a novel guided search approach for multi-dimensional optimization in the multi-sensor control command space, for minimization of a task-driven control objective function. It also utilizes Generalized Covariance Intersection (GCI) method for multi-sensor fusion. A step-by-step algorithm was detailed for SMC implementation of the proposed method with LMB filters running at each sensor node. 

Numerical studies were presented for several scenarios where numerous controllable (mobile) sensors track multiple moving targets with different levels of observability. The results demonstrated good performance in controlling numerous sensors (in terms of OSPA errors). They also showed that our proposed method runs substantially faster than the traditional exhaustive search-based technique. Indeed we showed that while the computational cost of traditional methods grow exponentially with increasing the number of sensors, our method has only second order computational complexity.

\section*{Acknowledgment}
This project was supported by the Australian Research Council through ARC Discovery grant~DP160104662, as well as National Nature Science Foundation of China grants~61673075.
%


\end{document}